\documentclass[reprint, superscriptaddress, amsmath, amssymb, aps, floatfix, 
prl]{revtex4-2}
\usepackage[latin9]{inputenc}
\usepackage{verbatim}
\usepackage{amsmath}
\usepackage{amssymb}
\usepackage{graphicx}

\usepackage{amsthm}
\usepackage{fancyhdr}
\usepackage{epsfig}
\usepackage{bbm}
\usepackage{subfigure}
\usepackage{float}
\usepackage{xcolor}
\usepackage[normalem]{ulem}
\usepackage[colorlinks=true,linkcolor=blue,citecolor=blue]{hyperref}
\usepackage{multirow}

\begin{document}
\title{Reversal of tracer advection and Hall drift in an interacting chiral fluid}

\author{Erik Kalz}
\thanks{These two authors contributed equally}
\affiliation{University of Potsdam, Institute of Physics and Astronomy, 
D-14476 Potsdam, Germany}

\author{Shashank Ravichandir}
\thanks{These two authors contributed equally}
\affiliation{Leibniz-Institute for Polymer Research, Institute Theory of 
Polymers, D-01069 Dresden, Germany}
\affiliation{Technical University of Dresden, Institute for Theoretical Physics, D-01069 Dresden, Germany}

\author{Johannes Birkenmeier}
\affiliation{University of Potsdam, Institute of Physics and Astronomy, 
D-14476 Potsdam, Germany}

\author{Ralf Metzler}
\thanks{Corresponding author}
\email{rmetzler@uni-potsdam.de}
\affiliation{University of Potsdam, Institute of Physics and Astronomy, 
D-14476 Potsdam, Germany}
\affiliation{Asia Pacific Centre for Theoretical Physics, KR-37673 Pohang, 
Republic of Korea} 

\author{Abhinav Sharma}
\thanks{Corresponding author}
\email{abhinav.sharma@uni-a.de}
\affiliation{Leibniz-Institute for Polymer Research, Institute Theory of 
Polymers, D-01069 Dresden, Germany}
\affiliation{University of Augsburg, Institute of Physics, D-86159 Augsburg, 
Germany}

\begin{abstract}
Chiral fluids are defined by broken mirror or time-reversal symmetry, 
giving rise to tensorial transport coefficients with antisymmetric components. 
A key example is the odd mobility tensor, which governs the response of a 
chiral tracer to an applied force and induces a characteristic transverse drift. 
While this response is well understood in the infinite dilution limit, the 
impact of interparticle interactions on the tracer dynamics remains largely 
unexplored. Here, we conduct an analytical and computational study of a chiral 
fluid with interparticle interactions and show that, under an external driving 
force, a chiral tracer can undergo a complete reversal of both its transverse 
Hall drift and its advection along the force. This reversal emerges from the 
interplay between odd mobility and interaction-mediated forces, resulting in a 
phenomenon reminiscent of absolute negative mobility.
\end{abstract}

\maketitle

\textit{Introduction.} 
Chiral fluids have garnered significant attention for their unconventional 
transport properties, with no counterparts in conventional fluids. 
These systems exhibit striking effects, such as edge flows and optimal transport 
occurring at boundaries \cite{nguyen2014emergent, perez2019bacteria,
yang2021topologically, caporusso2024phase} and 
interactions---even purely repulsive ones---enhancing rather than suppressing 
motion \cite{kalz2022collisions, reichhardt2022activerheology, ghimenti2023sampling, 
schick2024two}. The 
origin of these anomalous behaviors lies in the broken mirror or time-reversal 
symmetry inherent to chiral fluids. These unusual transport properties manifest
themselves 
across a diverse range of systems, from condensed matter settings like 
magnetic skyrmion structures \cite{troncoso2014brownian, 
reichhardt2015collective, buttner2018theory, gruber2023300}
and externally driven colloidal or macroscopic assemblies \cite{soni2019odd, 
massana2021arrested, bililign2022motile, grzybowski2000dynamic, wang2022order}
to biological systems 
featuring self-propelled chiral agents such as \textit{Escherichia coli} 
\cite{diluzio2005escherichia, lauga2006swimming, beppu2021edge}, 
\textit{Thiovulum majus} \cite{petroff2015fast},
\textit{Paenibacillus vortex} \cite{li2024robust}, or swimming starfish embryos 
\cite{tan2022odd}.
Further examples include vortex fluids 
\cite{riedel2005self, chen2024self}
and even granular systems 
such as strongly rotating or magnetized plasmas \cite{kahlert2012magnetizing,
shalchi2020perpendicular}.

A unifying framework has recently emerged to describe the transport properties 
of these diverse chiral systems using so-called \textit{odd} transport tensors. 
At the continuum level, chiral fluids are effectively characterized by concepts 
such as odd viscosity \cite{avron1998odd, banerjee2017odd, souslov2019topological,
fruchart2023odd, hosaka2023lorentz},
while on 
the discrete agent-based scale, their dynamics are described in terms of odd 
mobility \cite{reichhardt2019active, poggioli2023odd}
and odd diffusion 
\cite{hargus2021odd, kalz2022collisions, kalz2024oscillatory}. 
The defining feature of odd transport coefficients is their tensorial structure, 
which includes antisymmetric off-diagonal elements that directly arise from 
the fundamental symmetry principles of two-dimensional rotation 
\cite{bibnote_SUtwo} and encode the chiral nature of the fluid. Specifically, 
the odd mobility tensor is given by \cite{poggioli2023odd}
\begin{equation}
\label{definition_odd_mobility}
    \boldsymbol{\mu} = \mu_0 (\mathbf{1} + \kappa \boldsymbol{\epsilon}),
\end{equation}
where $\mathbf{1}$ denotes the identity tensor and $\boldsymbol{\epsilon}$ 
represents the totally antisymmetric Levi-Civita symbol in two dimensions. 
The mobility tensor relates the response velocity $\mathbf{v}$ of a tracer to an 
applied force $\mathbf{f}$ through $\mathbf{v} = \boldsymbol{\mu}\,  \mathbf{f}$. 
In Eq.~\eqref{definition_odd_mobility} $\mu_0$ is the bare mobility with 
dimensions $[\mu_0]=\mathrm{s}/\mathrm{kg}$ and $\kappa$ is the dimensionless 
characteristic \textit{odd-mobility} coefficient. 

In the absence of interactions, the response of a chiral tracer to an applied 
force follows directly from the odd mobility tensor. For a drift force 
$\mathbf{f} = f \hat{\mathbf{e}}_x$, the tracer moves with velocity $\mathbf{v} = 
\mu_0 f (1, -\kappa)^\mathrm{T}$, exhibiting a characteristic component 
transverse to the driving force. This motion, commonly associated with effects 
such as Hall viscosity \cite{avron1995viscosity, delacretaz2017transport, 
holder2019unified},
%berdyugin2019measuring
the skyrmion Hall effect 
\cite{litzius2017skyrmion, jiang2017direct}, or lift forces in chiral flows 
\cite{lier2023lift, cao2023memory}, is an intrinsic feature of odd mobility. 
While this behavior is well understood in the infinite dilution limit, the 
impact of interparticle interactions on the tracer's drift in a chiral fluid 
remains an open problem.
In this work, we systematically investigate this problem using analytical 
theory and simulations. By incorporating interactions within a chiral fluid, 
we reveal that a driven tracer can undergo a complete reversal of both its 
intrinsic Hall drift and its advection along the direction of the force---a 
striking effect reminiscent of absolute negative mobility 
\cite{eichhorn2002brownian, eichhorn2002paradoxical}.
In particular, we show that these effects emerge due 
to the interplay of interparticle interactions and their odd mobility, which 
govern how the driving force is mediated through the fluid. Importantly, our 
theoretical framework respects the fluctuation-dissipation relation (FDR), 
implying that these unconventional transport phenomena arise within equilibrium 
settings.

\textit{Model.} We model the chiral fluid as a system of $N+1$ interacting 
Brownian particles in two spatial dimensions. The chirality is encoded in the 
odd mobility $\boldsymbol{\mu}_i$ for particle $i\in\{0, \ldots, N\}$. Each 
particle is subjected to a constant external drift $\mathbf{f}_i$, such that 
the overdamped Langevin dynamics become
\begin{equation}
\label{odd_langevin_equation}
    \dot{\mathbf{r}}_i(t) = \boldsymbol{\mu}_i\mathbf{f}_i - \boldsymbol{\mu}_i\sum_{j\neq i}\nabla_i U(\mathbf{r}_i, \mathbf{r}_j), + \boldsymbol{\eta}_i(t),
\end{equation}
where $U$ is the two-body interaction potential, effecting in the interaction 
force $-\sum_{j\neq i}\nabla_i U(\mathbf{r}_i, \mathbf{r}_j)$ on particle $i$. 
Note that the response to both, the external drift and the interaction force has 
transverse components encoded in $\boldsymbol{\mu}_i$. $\boldsymbol{\eta}_i(t)$ 
is the driving Gaussian but nonwhite noise with mean 
$\langle \boldsymbol{\eta}_i(t)\rangle = \mathbf{0}$ and correlation $\langle 
\boldsymbol{\eta}_i(t)\boldsymbol{\eta}_j(t^\prime)\rangle = \mathbf{D}_i\, 
\delta_+(t - t^\prime)\delta_{ij} + \mathbf{D}_i^\mathrm{T}\, 
\delta_-(t - t^\prime) \delta_{ij}$ \cite{chun2018emergence, 
park2021thermodynamic}, where $\mathbf{D}_i=D_{0, i} (\mathbf{1} + \kappa_i 
\boldsymbol{\epsilon})$ is the odd diffusion of particle $i$, analogous to 
Eq.~\eqref{definition_odd_mobility}. Here $\delta_\pm(\cdot)$ represent 
variants of the Dirac delta distribution defined on the half-axis of the real 
numbers, $\mathbb{R}_\pm$, respectively \cite{bibnote_delta}. For analytical simplicity, we assume a 
tensorial FDR, i.e., $\mathbf{D}_i=\boldsymbol{\mu}_i\, k_\mathrm{B} T$, where 
$k_\mathrm{B}T$ is the thermal energy.
By tagging one particle, the tracer particle, we coarse-grain 
Eq.~\eqref{odd_langevin_equation} over the interactions with the host particles 
in the limit of low density $\phi\ll 1$ (see the Supplementary Material (SM) 
\cite{supplemental_material} for details). 
We adapt a geometric method in the coarsening~\cite{bruna2012diffusion, 
bruna2012excluded}. This allows us to analytically treat the oddness-affected 
probability fluxes, originating in the modified repulsive (so-called 
\textit{oblique} \cite{gilbarg2001elliptic}) boundary conditions introduced by 
the sterically interacting particles.
The drift-diffusion equation for the one-body probability density function 
(PDF) $p_t(\mathbf{x},t)$, to find the tracer in the interval $[\mathbf{x}, 
\mathbf{x}+ \mathrm{d}\mathbf{x}]$ at time $t$ becomes %(see also the End Matter)
\begin{equation}
\label{effective_FP_equation}
\frac{\partial p_t(\mathbf{x},t)}{\partial t} = \nabla 
\cdot\left[D \nabla - \mathbf{v}\right]p_t(\mathbf{x},t).
\end{equation} 
Here $D = \mathrm{diag} \left(\mathbf{D}_t \left[\mathbf{1} - 
4\phi \mathbf{\Gamma}\mathbf{D}_t\right]\right)$ is the effective diffusion 
coefficient, where $\mathbf{\Gamma} = 
(\mathbf{D}_t + \mathbf{D}_h)/\mathrm{det}(\mathbf{D}_t + \mathbf{D}_h)$. %$\mathbf{D}_t$ and $\mathbf{D}_h$ are the odd diffusion tensor of the tracer ($t$) and host particles ($h$), respectively. 
The tracer drift $\mathbf{v}$ is given by 
\begin{equation}
\label{direct_trans_mobility_tracer}
\mathbf{v} = \boldsymbol{\mu}_\mathrm{dir} \mathbf{f}_t + 
\boldsymbol{\mu}_\mathrm{indir} \mathbf{f}_h, 
\end{equation}
where the mobilities
\begin{align} 
\label{direct_mobility}
\boldsymbol{\mu}_\mathrm{dir} &=  \boldsymbol{\mu}_t - 4\phi
\mathbf{D}_t\boldsymbol{\Gamma}\boldsymbol{\mu}_t, \\
\label{transferred_mobility}
\boldsymbol{\mu}_\mathrm{indir} &= 4\phi
\mathbf{D}_t \boldsymbol{\Gamma} \boldsymbol{\mu}_h,
\end{align}
are referred to as the \textit{direct} and \textit{indirect} mobility of the 
tracer particle, respectively. Here subscripts $t$ and $h$ refer to tracer and 
host particles, respectively.

Both, $\boldsymbol{\mu}_\mathrm{dir}$ and $\boldsymbol{\mu}_\mathrm{indir}$ are 
influenced by interactions between the tracer and host particles. In an ordinary, 
nonchiral fluid, these mobilities correspond to distinct physical scenarios; 
when only the tracer particle is driven ($\mathbf{f}_h=\mathbf{0}$), 
$\boldsymbol{\mu}_\mathrm{dir}$ describes how the tracer advection is hindered 
by continuous collisions with the host particles at rest. Conversely, when the 
tracer remains at rest ($\mathbf{f}_t=\mathbf{0}$), while the host particles 
are driven, $\boldsymbol{\mu}_\mathrm{indir}$ quantifies how the tracer is 
advected by the 
host particle stream, an effect mediated by interparticle interactions. In both 
cases, the tracer drifts in the direction of the applied force 
\cite{hansen2013theory} (see also the SM 
\cite{supplemental_material}) a trend seen at all densities of the host 
particles. This intuitive behavior is reflected in the positivity of both 
transport coefficients, direct and indirect mobility. However, this is 
fundamentally different in a chiral fluid. We demonstrate that in a chiral 
fluid with interacting particles these mobilities can turn negative leading to 
reversed tracer drift.

\begin{figure*}
\centering
\includegraphics[width = 0.9\textwidth]{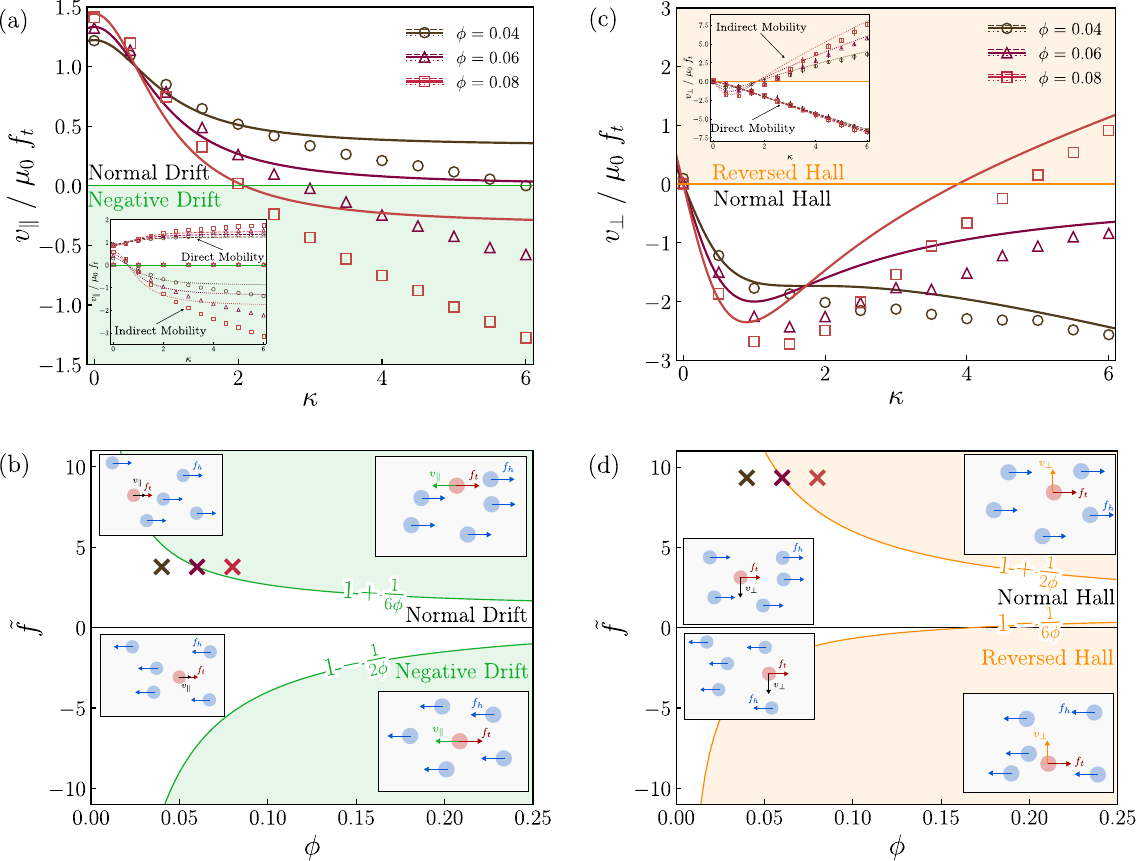}
\caption{\textit{Reversal of tracer drift.}
Components of the tracer drift $\mathbf{v}=(v_\parallel, v_\perp)$, 
Eq.~\eqref{direct_trans_mobility_tracer} (solid lines), of a driven chiral 
tracer in a driven chiral fluid. Symbols represent data from Brownian 
dynamics simulations. Error bars are smaller than the symbols. Tracer drift 
$\mathbf{f}_t = f_t\hat{\mathbf{e}}_x$ 
and host drift $\mathbf{f}_h = f_h\hat{\mathbf{e}}_x$ are taken to be 
parallel, the ratio of magnitudes $\tilde{f}=f_h/f_t$ serves as a control 
parameter. (a) Tracer tracer drift along the drift force $v_\parallel$ as 
a function of odd mobility $\kappa$ for $\tilde{f}=3.77$. The densities 
$\phi=0.04, 0.06, 0.08$ correspond to the three regions of the state 
diagram (b) of normal drift, asymptotically negative drift, and negative 
drift (indicated by crosses). 
(c) Hall-drift of the tracer $v_\perp$ as a function of odd mobility 
$\kappa$ for $\tilde{f}=9.33$. The densities $\phi=0.04, 0.06, 0.08$ 
correspond to the three regions of the state diagram (d) of ordinary 
Hall drift, asymptotic reversal, and reversal of Hall drift (indicated by 
crosses). Insets in (a) 
and (c) show the contribution of $\boldsymbol{\mu}_\mathrm{dir}$ (dashed 
lines) and $\boldsymbol{\mu}_\mathrm{indir}$ (dotted lines) according to 
Eqs.~\eqref{direct_mobility} and \eqref{transferred_mobility} to the 
negative drift and reversed Hall drift of the tracer. Sketches in (b) and 
(d) illustrate the tracer's drift in the respective region.}
\label{fig_drift_reversal}
\end{figure*}

\textit{Results.} We model the chiral fluid by fixing $\boldsymbol{\mu}_t= 
\boldsymbol{\mu}_h$ and $\mathbf{D}_t= \mathbf{D}_h$ and in particular $\kappa_t = 
\kappa_h \equiv \kappa \neq 0$. %As model parameters, we vary the dragging forces $f_t$ and $f_h \neq f_t$. These represent the minimal conditions necessary to reveal the nontrivial effects of interactions in chiral fluids. 
Despite this simplification, the model remains broadly applicable to describe 
systems with distinct particle responses, such as driven binary colloidal 
systems ~\cite{helbing2000freezing, dzubiella2002lane, poncet2017universal}, 
and particularly tracer particles in chiral fluids \cite{reichhardt2019active, 
yang2021topologically, reichhardt2022activerheology, poggioli2023odd}.
We first consider the case where the tracer 
particle and host particles are subjected to dragging forces parallel to 
each other along a direction, which we chose as the $x$ direction, i.e., 
$\mathbf{f}_t = f_t \hat{\mathbf{e}}_x$ and $\mathbf{f}_h = f_h 
\hat{\mathbf{e}}_x$. 
As model parameters, we vary the magnitudes of the dragging forces. The results 
are shown for a chosen $f_t$ and the ratio $\tilde{f} = f_h/f_t$.
The drift of the tracer is decomposed into its components along and 
perpendicular to the tracer 
force, which are denoted by $v_\parallel$ and $v_{\perp}$, respectively.

Consider the scenario in which $f_t,f_h > 0$, i.e., the chiral fluid and the 
tracer are dragged in the same direction. 
In contrast to the normal fluid, here the chiral tracer exhibits highly 
counterintuitive dynamics: for sufficiently large densities, the tracer's 
drift velocity along the applied force reverses direction despite $\tilde{f} >0$.
The tracer drift becomes negative when the oddness exceeds a threshold 
$\kappa> \kappa_\mathrm{rev}^{\parallel}(\tilde{f}, \phi)$ that depends 
both on $\tilde{f}$ and $\phi$, which need to obey $\tilde{f}> 1 + 1/(6\phi)$ 
(for analytical details, see the SM~\cite{supplemental_material}). The latter 
gives rise to a state diagram for negative tracer drift shown in
Fig.~\ref{fig_drift_reversal}(b) and we test the predictions against Brownian 
dynamics simulation results (see the SM \cite{supplemental_material} for details 
on simulations) for $\tilde{f}=3.77$ in Fig.~\ref{fig_drift_reversal}(a).
We observe three distinct regimes: at lower densities ($\phi=0.04$), the tracer 
exhibits ordinary drift, consistent with the 
applied force. At higher densities, in contrast, ($\phi=0.08$) the chiral tracer 
exhibits negative drift ($v_\parallel <0$).  At the threshold density 
$\phi=0.06$, the theory predicts an asymptotic negative drift as $\kappa \to \infty$. 
While the theory and simulations are in good agreement for small 
$\kappa$ ($\kappa \lesssim 2$), the measured effect is in fact much stronger 
than the theory predicts, compare Fig.~\ref{fig_drift_reversal}(a). 

To investigate the origin of the negative drift, we separately analyze the 
effects of direct and indirect mobility in Eq.~\eqref{direct_trans_mobility_tracer} 
by either switching off the host or tracer drift as shown in the inset of 
Fig.~\ref{fig_drift_reversal}(a).
The direct mobility, which remains positive for all $\kappa$ describes the 
normal drift response of the tracer in that it describes how the motion of a 
tracer is impeded due to interactions with host particles. The indirect mobility, 
in contrast, captures the enhancement of dynamics due to interactions with the 
chiral fluid. In other words, it is the response of the tracer to the driving 
of the host particles, which ultimately leads to a reversal of the tracer's 
drift. The theory predicts a second regime with negative drift 
($\tilde{f} < 1 - 1/(2\phi)$, i.e., $\tilde{f}<0$ in the dilute region); see 
Fig.~\ref{fig_drift_reversal}(b). Here, the tracer and host are dragged in 
opposite directions and thus, a reversal of tracer drift due to interactions 
might appear less surprising. 
More intriguingly, however, at sufficiently high oddness 
($\kappa>\kappa_{\mathrm{rev}}^\parallel(\tilde{f}, \phi)$) the tracer 
overcomes the transferred 
effect of interactions and again follows the dragging force, despite the 
opposing host-particle stream. 

We now focus on $v_{\perp}$, the tracer's drift velocity perpendicular to the 
applied force. From Eq.~\eqref{direct_trans_mobility_tracer}, we find that 
$v_{\perp} \propto - \kappa$ in the absence of interactions, which corresponds 
to the characteristic Hall drift of a chiral tracer. In chiral fluids, 
interactions with the host particles can lead to a similarly counterintuitive 
effect as for $v_\parallel$ and reverse the direction of $v_{\perp}$ despite 
$\tilde{f}>0$ and $\mathbf{f}_t \parallel \mathbf{f}_h$.
This is possible, if the chirality exceeds a threshold 
$\kappa>\kappa_\mathrm{rev}^\perp(\tilde{f}, \phi)$, that depends both 
on $\tilde{f}$ and $\phi$, which have to obey the relation 
$\tilde{f}> 1 + 1/(2\phi)$ (for analytical details, see the 
SM~\cite{supplemental_material}). The latter again gives rise to a state 
diagram for reversed Hall drift, shown in 
Fig.~\ref{fig_drift_reversal}(d). In Fig.~\ref{fig_drift_reversal}(c) we 
demonstrate that for a chosen value of $\tilde{f} = 9.33$, in a system with 
sufficiently high density, the chiral tracer reverses its intrinsic Hall drift. 
Again three distinct regimes are found; at low densities ($\phi=0.04$), the 
tracer particle exhibits a Hall-like drift, $v_{\perp} \propto -\kappa$. However, 
at large densities ($\phi=0.08$), the Hall drift becomes non-monotonic as a 
function of $\kappa$, ultimately resulting in a reversal of the Hall drift 
($v_\perp>0$). These regimes are separated by an asymptotically reversed Hall drift 
($\phi=0.06$). Analogously to the negative drift, the reversed Hall effect originates 
in indirect mobility, as can be seen in the inset in 
Fig.~\ref{fig_drift_reversal}(c). Similar to the negative drift, the reversal 
of Hall drift, thus, is induced 
by the indirect response of the tracer to the host driving in the dilute limit.
In fact, as evident from both state diagrams, 
Figs.~\ref{fig_drift_reversal}(b) and (d), a reversal of the Hall drift 
necessarily accompanies a negative drift within the current setup and cannot 
occur independently.

\begin{figure}
\begin{subfigure}
\centering
\includegraphics[width=0.95\linewidth]{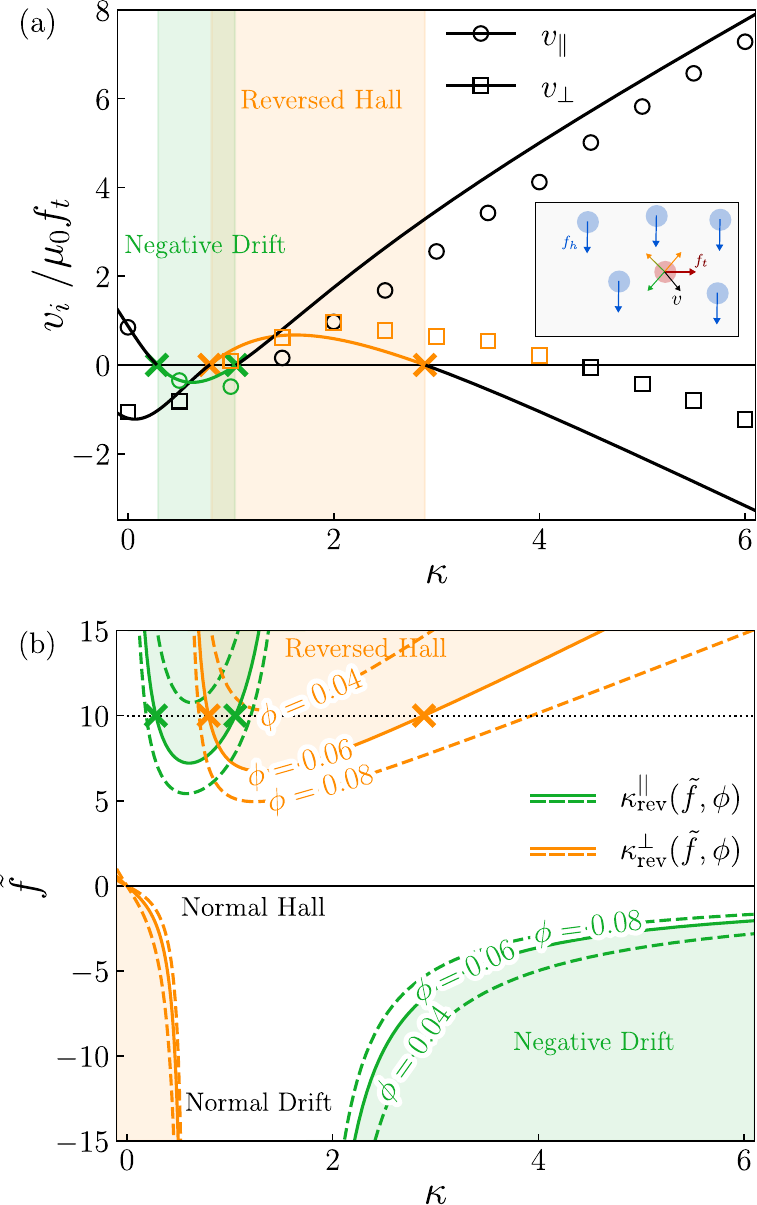}
\end{subfigure}
\caption{\textit{Orthogonal drift forces.} (a) Components $v_i/\mu_0 f_t$ of 
the tracer drift $\mathbf{v}=(v_\parallel, v_\perp)$ from 
Eq~\eqref{direct_trans_mobility_tracer} (solid lines), as a function of 
odd mobility $\kappa$ 
for the case of tracer drift $\mathbf{f}_t=f_t\hat{\mathbf{e}}_x$ perpendicular 
to host drift $\mathbf{f}_h=-f_h\hat{\mathbf{e}}_y$ in a system with density 
$\phi=0.06$. Symbols represent data from Brownian dynamics simulations. 
Error bars are smaller than the symbols. The ratio of magnitudes is chosen 
as $\tilde{f}=f_h/f_t = 10$ and 
is indicated in the theoretical state diagram (b) by the dotted line. The state 
diagram shows that for different densities $\phi=0.04, 0.06, 0.08$ the reversed 
Hall drift (orange) and negative drift (green) can be realized independently or 
simultanously. Crosses indicate the theoretically predicated on-, and offset of 
reversed drifts, and agree qualitatively well with simulations. Note that the 
choices of $\tilde{f}>0$ and $\alpha=\mathrm{arccos} (\mathbf{f}_t \cdot 
\mathbf{f}_h)=-\pi/2$ is equivalent to $\tilde{f}<0$ and $\alpha=\pi/2$ in (b).}  
\label{orthogonal_drift}
\end{figure}

The simultaneous reversal of both drift components is a direct consequence of 
the chosen setup, where the dragging forces are strictly parallel. 
A more general scenario is nonparallel drift forces, which have 
been thoroughly investigated, for example in driven colloidal systems 
\cite{dzubiella2002pattern, cividini2013diagonal},
in streamline flows \cite{shin2017accumulation} or in pedestrian dynamics 
\cite{helbing2005self, mullick2022analysis, bacik2023lane}.
With nonparallel forces, a broader range 
of tracer responses exists, allowing for control over its motion in all directions. 
For instance, Fig.~\ref{orthogonal_drift} presents the state diagram and 
simulation results for a scenario in which $\mathbf{f}_t$ is perpendicular to 
$\mathbf{f}_h$. In this case, the reversed Hall drift does not necessarily 
imply a negative drift, and both effects can be independently realized (see 
the SM for details~\cite{supplemental_material}). The parallel and perpendicular 
dragging cases illustrate the essential nontrivial behavior of chiral fluids and 
their tunability. However, our theory provides a general prediction for an 
arbitrary angle between the tracer and host forces. This fully general response 
is analyzed in the SM \cite{supplemental_material}.

\textit{Discussion.} 
We demonstrated that the transport properties of a tracer in a chiral fluid 
exhibit rich and nontrivial behavior, strongly shaped by interparticle 
interactions with the host fluid. Driven chiral tracers can undergo a reversal 
of their intrinsic Hall drift and experience negative drift, emerging from the 
subtle interplay of chirality, particle density, and drift magnitude.

Our findings have significant experimental and theoretical implications for the 
rheological properties of chiral fluids, highlighting the fundamental role 
of interaction effects in shaping their transport phenomena. These insights are 
directly relevant for tracer dynamics in skyrmionic systems 
\cite{zhang2023laminar, raab2024skyrmion}, 
or bacterial suspensions \cite{petroff2015fast, beppu2021edge, li2024robust},
where our predictions can be experimentally tested. Additionally, our results 
may have implications for the sedimentation of chiral tracers 
\cite{krapf2009chiral, huseby2025helical} 
with potentially different buoyancy than the (chiral) fluid 
\cite{herron1975sedimentation, singh2021bacterial, khain2022stokes}.
Moreover, these results open up new research perspectives on how chirality 
influences collective behaviors, such as laning transitions in mixtures of 
oppositely driven species \cite{helbing2000freezing, poncet2017universal, 
dzubiella2002lane}.
In these systems, spontaneous demixing 
occurs, leading to the formation of lanes with opposite flow. 

Finally, we remark on the intriguing resemblance between absolute negative 
mobility \cite{eichhorn2002brownian, eichhorn2002paradoxical}
and the negative drift observed 
in our system. This offers a fascinating example of a chiral fluid under flow 
acting as an engine, where the tracer, drifting opposite to the applied force, 
can perform work \cite{hanggi2009artificial}. The fact that this phenomenon can 
already occur in a passive system 
challenges conventional perspectives on nonequilibrium transport 
and highlights the rich, unexplored dynamics of 
chiral fluids. 

\textit{Acknowledgments.} E. K., A. S., and R. M. acknowledge support by the 
Deutsche Forschungsgemeinschaft (grants No. SPP 2332- 492009952, SH 1275/5-1 
and ME 1535/16-1). S. R. acknowledges support by Deutscher Akademischer 
Austauschdienst (grant No. 57693453).

\end{document}

\title{Supplementary Material: Reversal of tracer advection and Hall drift 
in an interacting chiral fluid}

\author{Erik Kalz}
\thanks{These two authors contributed equally}
\affiliation{University of Potsdam, Institute of Physics and Astronomy, 
D-14476 Potsdam, Germany}

\author{Shashank Ravichandir}
\thanks{These two authors contributed equally}
\affiliation{Leibniz-Institute for Polymer Research, Institute Theory of 
Polymers, D-01069 Dresden, Germany}
\affiliation{Technical University of Dresden, Institute for Theoretical Physics, D-01069 Dresden, Germany}

\author{Johannes Birkenmeier}
\affiliation{University of Potsdam, Institute of Physics and Astronomy, 
D-14476 Potsdam, Germany}

\author{Ralf Metzler}
\thanks{Corresponding author}
\email{rmetzler@uni-potsdam.de}
\affiliation{University of Potsdam, Institute of Physics and Astronomy, 
D-14476 Potsdam, Germany}
\affiliation{Asia Pacific Centre for Theoretical Physics, KR-37673 Pohang, 
Republic of Korea} 

\author{Abhinav Sharma}
\thanks{Corresponding author}
\email{abhinav.sharma@uni-a.de}
\affiliation{Leibniz-Institute for Polymer Research, Institute Theory of 
Polymers, D-01069 Dresden, Germany}
\affiliation{University of Augsburg, Institute of Physics, D-86159 Augsburg, 
Germany}

\maketitle
In this Supplementary Material, we provide the analytical and simulation 
background for the relations and figures shown in the main manuscript. In 
Section~\ref{sec:theory}, we present the theory for obtaining an effective 
drift-diffusion equation for a driven chiral tracer in a dilute suspension of 
driven chiral host particles. In Section~\ref{sec:setup} we 
present the setup of overdamped chiral particles, where the chirality is encoded 
in odd transport tensors. In Section~\ref{sec:interactions}, we outline the 
derivation of an interaction-corrected drift-diffusion equation and specify the 
model in Section~\ref{sec:tracer_host} to the tracer and host model for chiral 
particles.

In Section~\ref{sec:chiral_fluid}, we specify the generic model to the chiral 
fluid and present the relations to which we refer in the main manuscript. 
In Section~\ref{sec:parallel} we present the main setup of this 
study, parallel tracer and host drift forces and in Section~\ref{sec:orthogonal} 
the case of orthogonal drift forces. Section~\ref{sec:nonchiral} demonstrates 
that our model also captures other scenarios such as the nonchiral tracer 
particle in a chiral fluid, widely investigated in the literature, as well as 
that it
reduces to the ordinary nonchiral tracer in a nonchiral fluid as demonstrated 
in Section~\ref{sec:nonchiral_nonchiral}.

Finally, in Section~\ref{sec:simulation} we give details on the Brownian 
dynamics simulations.

\tableofcontents

\section{Modeling interacting chiral systems}
\label{sec:theory}

\subsection{Setup}
\label{sec:setup}

Chiral fluids under drift are characterized by a Hall effect, i.e., a drift 
causes a response in the transverse direction \cite{avron1995viscosity}. On 
the level of tracer dynamics, and in a two-dimensional system, this is 
encoded in the \textit{odd mobility} tensor \cite{reichhardt2022activerheology,
poggioli2023odd}, which is a second-rank isotropic transport tensor giving 
the tracer response on an applied force. It is given by 
\begin{equation}
\label{odd_mobility_matrix}
\boldsymbol{\mu} = \mu_0 (\mathbf{1} + \kappa\, \boldsymbol{\epsilon}),
\end{equation}
where $\mathbf{1}$ is the identity tensor and $\boldsymbol{\epsilon}$ is the 
fully antisymmetric Levi-Civita symbol in two dimensions ($\epsilon_{xx}= 
\epsilon_{yy}=0$ and $\epsilon_{xy} = -\epsilon_{yx}=1$). $\mu_0$ is the bare 
mobility with $[\mu_0] =\mathrm{s}/\mathrm{kg}$, and $\kappa$ is the 
characteristic odd mobility with $[\kappa] =1$. If we consider, as a 
demonstrative example, a drift force in $x$-direction, $\mathbf{f} = f 
\hat{\mathbf{e}}_x$, the average response velocity of a particle in the 
overdamped regime becomes $\langle \dot{\mathbf{r}}\rangle = \mu_0 f 
\left(\hat{\mathbf{e}}_x - \kappa \hat{\mathbf{e}}_y\right)$. Thus, odd mobility 
encodes the characteristic Hall-like response ($\propto - \kappa 
\hat{\mathbf{e}}_y$) of a chiral particle. A complete description of the 
overdamped chiral dynamics requires to specify the diffusion coefficient. We, 
therefore, assume that the system obeys a fluctuation-dissipation relation (FDR) 
of the form $\mathbf{D}=k_\mathrm{B}T\, \boldsymbol{\mu}$. Such behavior 
is known as \textit{odd-diffusion} in the literature \cite{hargus2021odd,
kalz2022collisions,kalz2024oscillatory} and characterizes the broken 
time-reversal symmetry of a chiral overdamped fluid. $k_\mathrm{B}T$ here 
is the thermal energy with $k_\mathrm{B}$ as the Boltzmann constant and $T$ the 
temperature of the medium.

We specifically want to investigate the effects of inter-particle interactions 
on the transport properties of the chiral fluid. We, therefore, model the 
chiral fluid as an isotropic system of $N$ interacting Brownian particles 
in two spatial dimensions. The overdamped Langevin equation for the position 
$\mathbf{r}_i(t) \in \mathbb{R}^2$ of particle $i\in \{1, \ldots, N\}$ takes 
the form
%diffusing in a bounded domain $\Omega \subset \mathbb{R}^2$ 
\begin{equation}
    \label{overdamped_Langevin_eq}
    \dot{\mathbf{r}}_i(t) = \boldsymbol{\mu}_i\, \mathbf{F}_i(\vec{\mathbf{r}}) + \boldsymbol{\eta}_i(t).
\end{equation}
Here $\mathbf{F}_i(\vec{\mathbf{r}}) = \mathbf{f}_i - \nabla_i U_{N}
(\vec{\mathbf{r}})$ is the total force acting on particle $i$. It is composed of 
an external force $\mathbf{f}_i(\mathbf{r}_i)$ acting on the position of 
particle $i$ and the interaction force $- \nabla_i U_{N}(\vec{\mathbf{r}})$ that 
stems from a generic $N$-body interaction potential $U_{N}(\vec{\mathbf{r}})$. 
$\vec{\mathbf{r}} = (\mathbf{r}_1(t), \ldots, \mathbf{r}_N(t))$ is the vector 
of all particle coordinates and $\nabla_i$ the partial differential operator 
for the position of particle $i$. Each chiral particle $i$ responds on the total 
force $\mathbf{F}_i(\vec{\mathbf{r}})$ with an odd mobility of the form of 
Eq.~\eqref{odd_mobility_matrix}, $\boldsymbol{\mu}_i=\mu_0^{(i)}(\mathbf{1} + 
\kappa_i \boldsymbol{\epsilon})$.

$\boldsymbol{\eta}_i(t)$ in Eq.~\eqref{overdamped_Langevin_eq} is a 
Gaussian noise, which has mean zero, $\langle \boldsymbol{\eta}_i(t)\rangle = 
\boldsymbol{0}$, and a nonwhite correlation $\langle \boldsymbol{\eta}_i(t)
\otimes\boldsymbol{\eta}_j(t^\prime)\rangle = \mathbf{D}_i\, \delta_+
(t - t^\prime)\delta_{ij} + \mathbf{D}_i^\mathrm{T}\, \delta_-(t - t^\prime) 
\delta_{ij}$ \cite{chun2018emergence,park2021thermodynamic}. $\mathbf{D}_i = 
D_0^{(i)}(\mathbf{1} + \kappa_i \boldsymbol{\epsilon})$ here is the odd 
diffusion tensor of particle $i$ and $(\cdot)^\mathrm{T}$ denotes a matrix 
transpose. $\delta_\pm(\cdot)$ represent variants of the Dirac delta distribution 
defined on the half-axis of the real numbers, i.e., $\int_0^\infty\mathrm{d}u\, 
\delta_+(u) = \int_{-\infty}^0\mathrm{d}u\, \delta_-(u) = 1$ and 
$\int_{-\infty}^0\mathrm{d}u\, \delta_+(u) = \int_0^\infty\mathrm{d}u\, 
\delta_-(u) = 0$. Further, $\otimes$ denotes the outer product, $(\mathbf{a} 
\otimes \mathbf{b})_{\alpha\beta} = a_\alpha b_\beta$ for some vectors 
$\mathbf{a}$ and $\mathbf{b}$ and Greek indices denoting components. 

The non-white noise correlation implies that a corresponding probabilistic 
description to the Langevin dynamics in Eq.~\eqref{overdamped_Langevin_eq} 
cannot be derived using standard techniques \cite{gardiner2009handbook,
risken1989fokker}. Using methods from (stochastic) functional calculus 
\cite{hanggi1978correlation, hanggi1995colored}, however, it is possible to 
derive the time-evolution equation 
for the probability density function (PDF), which also yields the correct 
probability flux, characterized by chirality-induced nonzero rotation as 
demonstrated in Refs.~\cite{chun2018emergence, park2021thermodynamic}. 
The correct time-evolution equation for the $N$-body joint PDF, 
$\mathcal{P}_N(t) = \mathcal{P}_N(\vec{\mathbf{x}},t) = \left\langle 
\prod_{i=0}^N\delta(\mathbf{r}_i(t) - \mathbf{x}_i)\right
\rangle_{\vec{\boldsymbol{\eta}}}$ 
to find each particle $i$ within $[\mathbf{x}_i, \mathbf{x}_i + 
\mathrm{d}\mathbf{x}_i]$ at time $t$ takes the form of $\partial 
\mathcal{P}_N(t)/ \partial t = -\sum_{i=1}^N \nabla_i \cdot 
\boldsymbol{\mathcal{J}}_i(\vec{\mathbf{x}}, t)$, where the diffusive flux 
$\boldsymbol{\mathcal{J}}_i(\vec{\mathbf{x}}, t)$ is given as 
\begin{align}
\label{many_body_flux}
\boldsymbol{\mathcal{J}}_i(\vec{\mathbf{x}}, t) &= - [\mathbf{D}_i \nabla_i + 
\boldsymbol{\mu}_i \nabla_i U_N(\vec{\mathbf{x}})  - \boldsymbol{\mu}_i 
\mathbf{f}_i]\, P_N(\vec{\mathbf{x}},t).
\end{align}
and thus has the characteristic, chirality-induced nonzero rotation, 
$\nabla_i \wedge \boldsymbol{\mathcal{J}}_i \propto \kappa_i$, particularly in 
the diffusive part.

\subsection{Interactions in the dilute limit}
\label{sec:interactions}

We aim to derive the effect of interparticle interactions on the dynamical 
behavior of chiral particles. To gain analytical insight we restrict our 
analysis to the dilute limit, where the potential energy is assumed to be the 
sum of radial pair potentials $U$, i.e., $U_{N}(\mathbf{x},t) = 
\sum_{i,j=1, i\neq j}^{N} U(x_{ij})$, and $x_{ij}=|\mathbf{x}_i - \mathbf{x}_j|$. 
We further assume the interaction to be of the hard-disk nature, i.e., we 
assume an excluded volume condition for the particles of diameter $d$, i.e., 
$U(x_{ij})= 0$ for $x_{ij} \geq d$ or $U(x_{ij})=\infty$ for $x_{ij} < d$ for 
any two particles $i$ and $j$, $i\neq j$. 

The time-evolution equation for the joint two-body PDF $\mathcal{P}_2(t) = 
\mathcal{P}_2(\mathbf{x}_1, \mathbf{x}_2, t)$ can be specified from 
Eq.~\eqref{many_body_flux} to be
\begin{align}
    \label{FPE_on_whole_space}
    &\frac{\partial}{\partial t} \mathcal{P}_2(t) = \nabla_1 \cdot 
    \left[\mathbf{D}_1 
    \nabla_1 - \boldsymbol{\mu}_{1}\left(\mathbf{f}_1-\nabla_1 U(x_{12}) 
    \right)\right] \mathcal{P}_2(t) \nonumber\\
    & \quad + \nabla_2 \cdot \left[\mathbf{D}_2 \nabla_2 + 
    \boldsymbol{\mu}_{2}\left(\mathbf{f}_2-\nabla_2 U(x_{12}) \right)\right] 
    \mathcal{P}_2(t). 
\end{align}
We are interested in deriving an effective time-evolution equation for the
one-body PDF, $p_1(\mathbf{x}_1,t)$, of particle one, exemplarily, by integrating 
out particle two in Eq.~\eqref{FPE_on_whole_space}. Formally, 
\begin{align}
    p_1(\mathbf{x}_1,t) &= \int\mathrm{d}\mathbf{x}_2\ 
    \mathcal{P}_2(\mathbf{x}_1, \mathbf{x}_2, t) \nonumber \\
    & = \int_{\mathbb{R}^2 \setminus  \mathrm{B}_d(\mathbf{x}_1)}\mathrm{d}
    \mathbf{x}_2\ 
    P_2(\mathbf{x}_1, \mathbf{x}_2, t).
\end{align}
Here we took advantage of the excluded volume condition of the hard-disk 
interaction by defining the two-particle joint PDF as $\mathcal{P}_2(\mathbf{x}_1, 
\mathbf{x}_2,t) = \Theta(x_{12} - d)\, P_2(\mathbf{x}_1, \mathbf{x}_2,t)$, 
where $\Theta(\cdot)$ is the (left-sided) Heaviside step function. 
$\mathbb{R}^2 \setminus \mathrm{B}_d(\mathbf{x}_i)$ here denotes the reduced 
configuration space for particle two, given that 
particle one is at position $\mathbf{x}_1$ and created the excluded volume of 
$\mathrm{B}_d(\mathbf{x}_1)$. This reduced configuration space makes the 
integration over particle two nontrivial, as it creates an (inner) moving boundary 
that must be respected in the coarse-graining procedure. Originating in the 
excluded volume condition of the hard disks, the non-overlap condition of the 
hard-disk interaction translates into a no-flux condition for the 
probabilistic flux. Here the special nature of chirality has to be 
respected, as this extends the usual Neumann-boundary condition into a 
so-called oblique boundary condition \cite{gilbarg2001elliptic}. This allows 
for an orthogonal flux, the mathematical representation of chirality of the 
particles. 

The effective time-evolution equation for $p_1(t) = p_1(\mathbf{x}_1,t)$ 
becomes \cite{bruna2012diffusion, bruna2012excluded, kalz2022diffusion, 
kalz2022collisions}
\begin{equation}
    \label{effective_fpe}
    \frac{\partial}{\partial t} p_1(t) = \nabla_1 \cdot [\mathbf{D}_1\nabla_1 
    - \boldsymbol{\mu}_{1}\, \mathbf{f}_1]\, p_1(t) + 
    I_{12}(\mathbf{x}_1,t),
\end{equation}
where $I_{12}(\mathbf{x}_1,t)$ encodes the two-body interaction contribution 
and is given by
\begin{align}
    \label{collision_integral}
    I_{12}(\mathbf{x}_1,t) &= -\int_{\partial\mathrm{B}_d(\mathbf{x}_1)}
    \mathrm{d}S_2\ \mathbf{n}_2 \cdot [\mathbf{D}_1^\mathrm{T}\nabla_1 + 
    \mathbf{D}_1\nabla_2]\, P_2(t).
\end{align}
Here $\mathrm{d}S_2\, \mathbf{n}_2$ is the directed surface element of 
particle two.

To approximate $I_{12}$ systematically in the two-body limit, we expand $P_2(t)$ 
as a perturbation series in $\sigma = d/L$, where $L$ is a typical length 
scale of the domain of diffusion, i.e., $P_2(t) = P_2^{(0)}(t) + \sigma 
P_2^{(1)}(t) + \mathcal{O}(\sigma^2)$ and insert that into 
Eq.~\eqref{collision_integral}. Given the local nature of the interaction 
between particles in space, the key idea is to divide space into regions where 
particles are either well-separated ($|\mathbf{x}_1 - \mathbf{x}_2| \gg \sigma$) 
or closely interacting ($|\mathbf{x}_1 - \mathbf{x}_2| \sim \sigma$). $I_{12}$ 
is expected only to contribute in these interaction regions. Solutions in both 
regions are then matched using the method of asymptotic expansions 
\cite{bender1999advanced}. Within the two-particle picture $I_{12}$ can be 
evaluated exactly and becomes \cite{kalz2022diffusion}
\begin{align}
    \label{evaluated_collision_integral_12}
    &I_{12}(\mathbf{x}_1,t) = \pi\sigma^2\, \nabla_{1} \cdot 
   \mathbf{D}_1\left[ \left(\mathbf{1} + \boldsymbol{\Gamma}\mathbf{D}_2
    \right)\, p_1 \nabla_{1} p_2 
      \right. \nonumber\\
&\quad  \left.-\boldsymbol{\Gamma}\, \mathbf{D}_1\, p_2 \nabla_{1} p_1 + \boldsymbol{\Gamma} \left(\boldsymbol{\mu}_{1}\, 
    \mathbf{f}_1 - \boldsymbol{\mu}_{2}\, 
    \mathbf{f}_2\right)p_1 p_2 \right],
\end{align}
where we have defined $\boldsymbol{\Gamma} = (\mathbf{D}_1 + \mathbf{D}_2)/
\mathrm{det}  (\mathbf{D}_1 + \mathbf{D}_2)$, and abbreviated $p_1 = 
p_1(\mathbf{x}_1,t),\ p_2 = p_2(\mathbf{x}_1,t)$. Note that $P_2^{(0)}(t)$ does 
not contribute to $I_{12}$, as it represents the interaction-free contribution 
of the perturbation problem. 

In the dilute limit, this model now can be extended to contain $N_1$ particles 
of species one and $N_2$ of species two with $(N_1 + N_2)\sigma 
\ll 1$. $N_2\, I_{12}$ thus represents the interspecies interaction contribution 
to Eq.~\eqref{effective_fpe}. However, for a tracer particle of species one, 
there are also $(N_1-1)$ intra-species interactions, which are not accounted 
for by $I_{12}$ yet. They can be, however, included by considering a fictitious 
analog of Eq.~\eqref{effective_fpe}, where all particle labels are replaced by 
those of particle 
one. It is thus possible, to derive the intra-species correction as 
$(N_1-1)\, I_{11}$. The time-evolution equation for $p_1(\mathbf{x}, t)$ thus 
contains the interaction-free term, 
Eq.~\eqref{effective_fpe}, inter-species interaction contributions, 
Eq.~\eqref{evaluated_collision_integral_12}, and intra-species contributions. 

\subsection{Tracer and host model}
\label{sec:tracer_host}

We specify the model to that of a tracer and a host species, i.e. we assign 
one species to describe the tracer particle, $N_1 =1$, and the other species 
to describe the host particles $N_2 =N$, such 
that we have $N_1 + N_2 = N + 1$ particles in total. We denote the tracer 
species by the subscript $t$ and the host species 
by the subscript $h$, i.e., $p_1(t) \to p_t(t)$ and $p_2(t) \to p_h(t)$. We 
define where $c=\pi\sigma^2/4$ as the area of a particle and 
$\phi(\mathbf{x}, t) = N c\ p_h(\mathbf{x},t)$ as the area concentration of 
the host species. This, for large $N$, approximately equals the full area 
concentration of particles in the system. 

We assume $\phi(\mathbf{x}, t) = \phi$ as constant, and the diluteness 
assumption implies $\phi\ll 1$. The equation for $p_t(t)$ can be brought into 
the form of a drift-diffusion equation for the tracer, which is reported in the 
main manuscript and reads
\begin{equation}
\label{effective_FP_equation}
\frac{\partial p_t(\mathbf{x},t)}{\partial t} = \nabla 
\cdot\left[D \nabla - \mathbf{v}(\mathbf{x})
\right]p_t(\mathbf{x},t),
\end{equation} 
where
\begin{align}
D &= \mathrm{diag} \mathbf{D}_t\left[ \mathbf{1}- 
4\phi\, \mathbf{\Gamma}\mathbf{D}_t \right], \\
\label{general_tracer_drift}
\mathbf{v}(\mathbf{x}) &= \boldsymbol{\mu}_t\mathbf{f}_t(\mathbf{x})  - 4\phi\, 
\mathbf{D}_t \mathbf{\Gamma} \left(\boldsymbol{\mu}_t \mathbf{f}_t -
\boldsymbol{\mu}_h \mathbf{f}_h\right),
\end{align}
are the diffusion coefficient and tracer drift, respectively. 
\section{The driven chiral fluid}
\label{sec:chiral_fluid}

\begin{figure}
\centering
\includegraphics[width=1\linewidth]{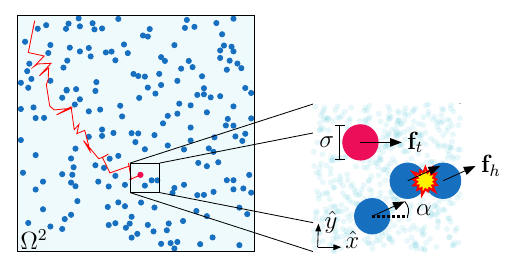}
\caption{\textit{Sketch.} Schematic of a driven chiral fluid, where tracer 
(red) and host (blue) particles experience the drift forces $\mathbf{f}_t$ and 
$\mathbf{f}_h$, respectively. The drift forces can be non-parallel in general, 
i.e., $\alpha= \mathrm{arccos}(\mathbf{f}_t \cdot \mathbf{f}_h) \in 
[-\pi/2, \pi/2)$. The tracer particle is experiencing the ordinary Hall 
drift $(\mathbf{v}_\mathrm{eff})_y \propto - \kappa$, which is curving the 
trajectory.}
    \label{fig:setup_chiral_drift}
\end{figure}

\begin{figure}
\begin{subfigure}\centering
\includegraphics[width=0.95\linewidth]{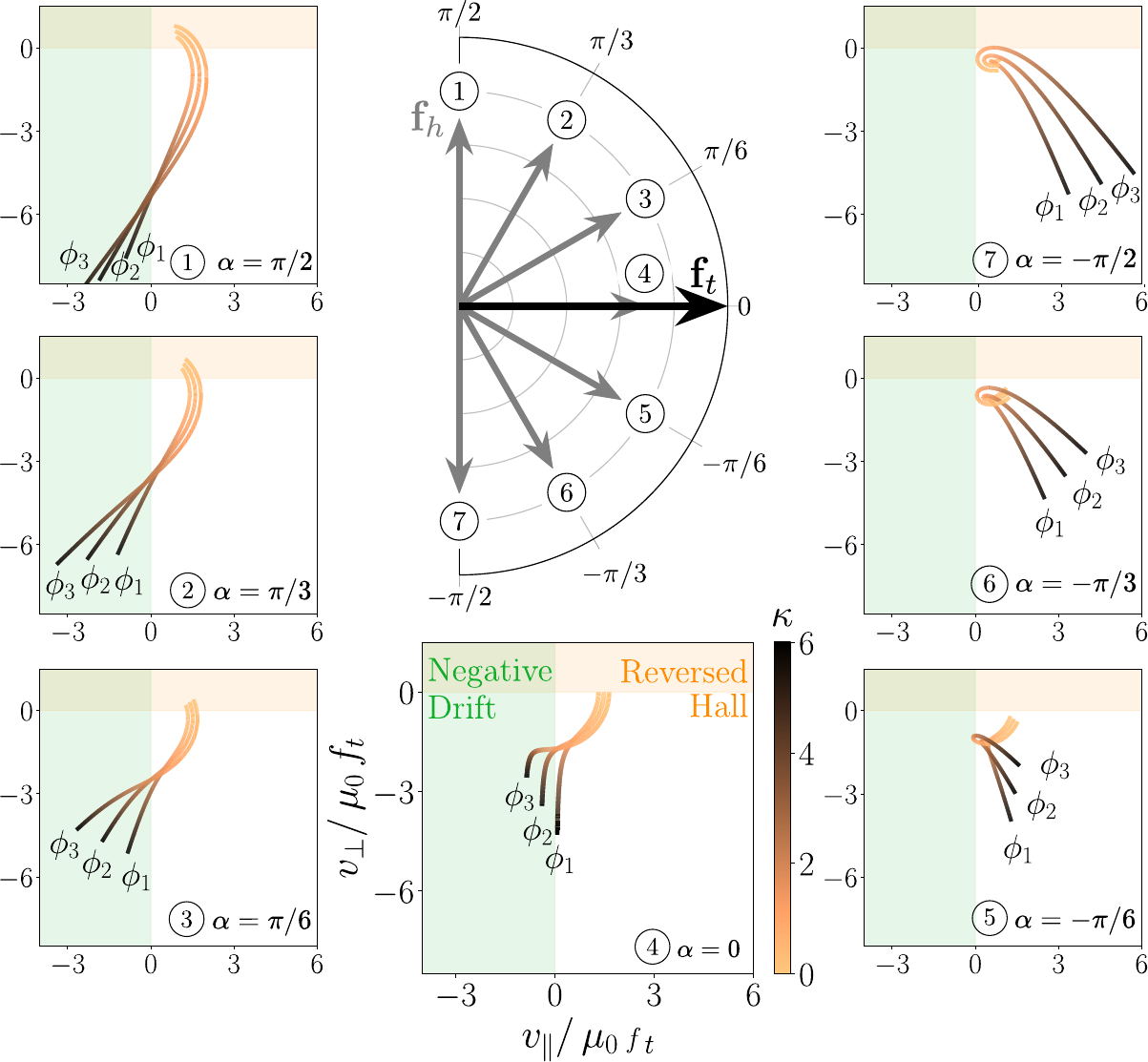}
\end{subfigure}
\caption{\textit{Nonparallel drift forces.} Each data point corresponds to the 
parallel and perpendicular drift $(v_\parallel, v_\perp)/\mu_0 f_t$ of the 
chiral tracer in a chiral host medium. For a given density and angle 
$\alpha= \mathrm{arccos}(\mathbf{f}_t \cdot \mathbf{f}_h)$ between the tracer 
and host force, the curves show the variation of the drift velocities with 
varying $\kappa$. The tracer response becomes highly tunable. Shown are three 
different densities $\phi_1 =0.04,\ \phi_2=0.06$ and $\phi_3=0.08$ for a 
ratio of drift magnitudes $\tilde{f}=f_h/f_t =5$. The situation $\tilde{f}>0$ 
and $\alpha \in [-\pi/2, \pi/2]$ is equivalent to $\tilde{f}<0$ and 
$\alpha\in [\pi/2, -\pi/2]$.}
\label{all_directional_drift}
\end{figure}

We model a chiral fluid of all-identical chiral particles by 
$\boldsymbol{\mu}_t= \boldsymbol{\mu}_h$ and $\mathbf{D}_t= \mathbf{D}_h$ and 
in particular $\kappa_t = \kappa_h \equiv \kappa \neq 0$. Tracer and host 
particles are subjected to a drift, $\mathbf{f}_t,\ \mathbf{f}_h$, respectively. 
If we were to consider equal drift of host and tracer particles, i.e., 
$\mathbf{f}_t = \mathbf{f}_h$, it can be easily seen that 
Eq.~\eqref{effective_FP_equation} is Galileian invariant under the transformation 
$\mathbf{x} \to \mathbf{x} - \mathbf{v}t$ and it becomes a pure 
(interaction-corrected) diffusion equation. To gain insight into the drift 
properties of a chiral tracer, it is, therefore, meaningful to consider a 
situation where $\mathbf{f}_t \neq \mathbf{f}_h$. We consider a situation of 
spatially constant drift and choose the coordinate system such that 
$\mathbf{f}_t = f_t \hat{\mathbf{e}}_x$. The host-driving, thus, takes the 
general form of $\mathbf{f}_h = f_h (\cos(\alpha), \sin(\alpha))^\mathrm{T}$, 
$\alpha \in [-\pi/2, \pi/2)$. The ratio of drift magnitudes $\tilde{f} = 
f_h/f_t$ serves as a control parameter. The tracer drift in 
Eq.~\eqref{general_tracer_drift} is decomposed along and perpendicular to the 
tracer force and is denoted by $v_\parallel$ and $v_{\perp}$, respectively. 
It becomes
\begin{widetext}
\begin{equation}
\label{tracer_drift_with_angle}
\begin{pmatrix}v_\parallel\\ v_\perp \end{pmatrix} = 
\mu_0 f_t \left[ \begin{pmatrix}1 \\ -\kappa\end{pmatrix} 
 - \frac{2\phi}{1 + \kappa^2} \begin{pmatrix}
     (1 - 3\kappa^2)(1 - \tilde{f}\cos(\alpha)) + \tilde{f}\sin(\alpha)\, 
     \kappa(\kappa^2 - 3) \\ \kappa(\kappa^2 - 3) (1 - \tilde{f}\cos(\alpha)) - 
     \tilde{f}\sin(\alpha)(1 - 3\kappa^2)
 \end{pmatrix}
\right].
\end{equation}
\end{widetext}
An illustration of the setup can be seen in Fig.~\ref{fig:setup_chiral_drift}, 
and the rich interplay of chirality, density and drift forces for the tracer 
response can be seen in Fiq.~\ref{all_directional_drift}. 

\subsection{Parallel drift forces}
\label{sec:parallel}

In the main manuscript, we prominently consider the situation of 
$\mathbf{f}_t \parallel \mathbf{f}_h$, i.e., $\alpha=0$. The tracer drift of 
Eq.~\eqref{tracer_drift_with_angle} specifies to 
\begin{equation}
    \begin{pmatrix}v_\parallel\\ v_\perp \end{pmatrix} = 
\mu_0 f_t \left[ \begin{pmatrix}1 \\ -\kappa\end{pmatrix} 
 - \frac{2\phi}{1 + \kappa^2} (1 - \tilde{f}) \begin{pmatrix}
     1 - 3\kappa^2 \\ \kappa(\kappa^2 - 3) \end{pmatrix}\right].
\end{equation}

Negative tracer drift, $v_\parallel/(\mu_0f_t)<0$, becomes possible for 
$\tilde{f}>1 + 1/(6\phi)$ and $\kappa > \kappa_\mathrm{rev}^{\parallel}(\tilde{f}, 
\phi)$, or $\tilde{f}<1 - 1/(2\phi)$ and $0<\kappa < 
\kappa_\mathrm{rev}^{\parallel}(\tilde{f}, \phi)$, where 
\begin{equation}
\kappa_\mathrm{rev}^{\parallel}(\tilde{f}, \phi) =  \sqrt{\frac{1 + 
2\phi(\tilde{f} -1)}{6\phi(\tilde{f}-1)-1}}. 
\end{equation}
Note that the second condition $\tilde{f}<1 - 1/(2\phi)$ implies $\tilde{f}<0$ 
for $\phi\in[0, 0.5)$, the regime of validity of the model 
\cite{bruna2012diffusion}, implying that tracer and host particles are driven 
in opposite directions. %Further, the adcvection along the tracer is invariant under $\kappa \to -\kappa$ and the negative drift occurs at $\kappa < - \kappa_\mathrm{rev}^{\parallel}(\tilde{f}, \phi)$.  

A reversal of the intrinsic Hall drift of the tracer, $v_\perp/(\mu_0f_t)>0$, 
becomes possible for $\tilde{f}>1 + 1/(2\phi)$ and $\kappa > 
\kappa_\mathrm{rev}^{\perp}(\tilde{f}, \phi)$, or $\tilde{f}<1 - 1/(6\phi)$ 
and $0<\kappa < \kappa_\mathrm{rev}^{\perp}(\tilde{f}, \phi)$, where 
\begin{equation}
\kappa_\mathrm{rev}^{\perp}(\tilde{f}, \phi) =  \sqrt{\frac{1 + 
6\phi(\tilde{f} -1)}{2\phi(\tilde{f}-1)-1}}. 
\end{equation}
Note that the second condition $\tilde{f}<1 - 1/(6\phi)$ implies $\tilde{f}<0$ 
for $\phi\in[0, 0.17]$, implying an opposite drift of tracer and host particles 
in the dilute regime. 

\begin{figure}
\centering
\includegraphics[width=\linewidth]{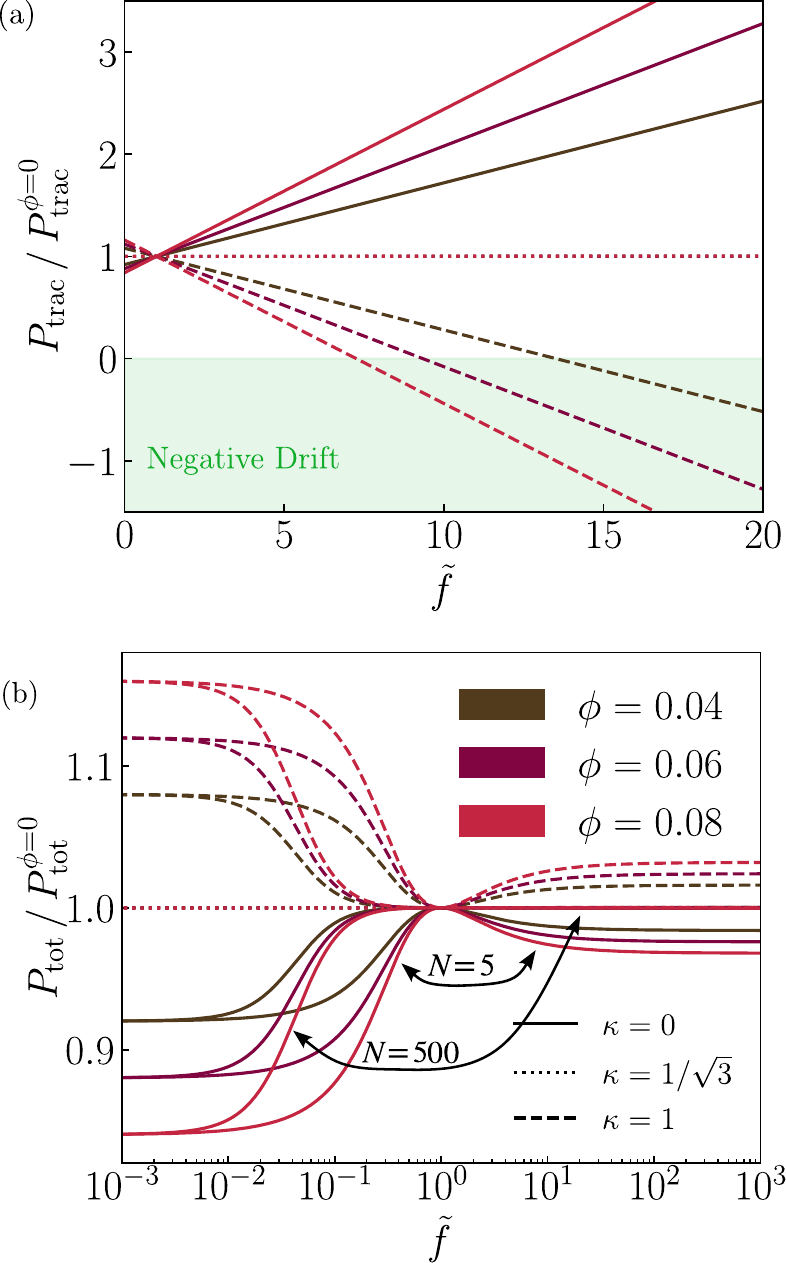}
\caption{\textit{Power.} (a) Tracer power according to Eq.~\eqref{tracer_power}. 
(b) Total power according to Eq.~\eqref{total_power} for two different numbers 
of total particles $N+1=6$ and $N+1=501$. Both (a) and (b) share the same 
legends which coincides with the choices for $\phi=0.04, 0.06, 0.08$ of the 
main manuscript figures.}
\label{fig:power}
\end{figure}

As stated in the main manuscript, the negative tracer drift, 
$v_\parallel/(\mu_0f_t)<0$, amounts to an interpretation of absolute negative 
mobility (ANM) of the tracer. This interpretation can be strengthened by 
analyzing the power it takes to drag the tracer. The power is defined as 
$P_\mathrm{trac} = \mathbf{v} \cdot \mathbf{f}_t$. In the system under 
consideration, it is given by 
\begin{equation}
\label{tracer_power}
    \frac{P_{\mathrm{trac}}}{P_\mathrm{trac}^{\phi=0}} = 1 - 2\phi\, 
    (1 - \tilde{f}) \frac{1 - 3\kappa^2}{1 + \kappa^2},
\end{equation}
normalized by the power to drag the tracer in the absence of interactions 
$P_\mathrm{trac}^{\phi=0} = \mu_0\, f_t^2$. For 
$\kappa>\kappa_\mathrm{rev}^{\parallel}(\tilde{f}, \phi)$ and sufficiently 
large $\tilde{f}$, indeed $P_\mathrm{trac}<0$ suggesting that the tracer can 
perform work against the drift force and thus resembles ANM, as shown in 
Fig.~\ref{fig:power} (a). However, the negative tracer drift originates entirely 
from the interaction transferred response to the host driving and disappears 
in the limit of vanishing host drift $\tilde{f} = f_h/f_t \to 0$, as illuminated 
in the main manuscript. The distinction to ANM becomes even clearer when 
considering the total power to drag all particles $P_{\mathrm{tot}} = 
\mathbf{v} \cdot \mathbf{f}_t + N \mathbf{v}_h \cdot \mathbf{f}_h$, which further 
includes the effective drift of the $N$ host particles, $\mathbf{v}_h$. The 
latter can be obtained from an analog to Eq.~\eqref{effective_FP_equation}. The 
total power reads
\begin{equation}
\label{total_power}
    \frac{P_{\mathrm{tot}}}{P_\mathrm{tot}^{\phi=0}} = 1 - 2\phi \frac{(1 - \tilde{f})^2}{1 + N\tilde{f}^2}\, \frac{1 - 3\kappa^2}{1 + \kappa^2},
\end{equation}
where $P_\mathrm{tot}^{\phi=0}= \mu_0 (f_t^2 + N f_h^2)$ again is the total 
power in the absence of interactions. The relevant term 
$(1 -\tilde{f})^2/(1 + N\tilde{f}^2)$ asymptotically approaches $1$ for 
$f_t\gg f_h$ and $1/N \to 0$ for $f_t\ll f_h$ and large $N$. Consequently, 
in the dilute limit ($\phi\ll 1$), $P_{\mathrm{tot}}$ remains strictly positive 
as shown in Fig.\ref{fig:power} (b) and in agreement with the second law of 
thermodynamics \cite{eichhorn2002paradoxical,eichhorn2002brownian}. 
Particularly $\kappa>0$ only increases the total power to drag the particles.

\subsection{Orthogonal drift forces}
\label{sec:orthogonal}

In the main manuscript, we further consider the situation of orthogonal drift 
forces, i.e., $\alpha = \pm \pi/2$. The tracer drift of 
Eq.~\eqref{tracer_drift_with_angle} specifies to 
\begin{align}
    \begin{pmatrix}v_\parallel\\ v_\perp \end{pmatrix} &= 
\mu_0 f_t \left[ \begin{pmatrix}1 \\ -\kappa\end{pmatrix} \right. \nonumber 
\\ &\quad \left.
 - \frac{2\phi}{1 + \kappa^2} \begin{pmatrix}
     1 - 3\kappa^2  - |\tilde{f}|(\kappa^3 - 3\kappa)\\ \kappa^3 - 3\kappa + 
     |\tilde{f}|(3\kappa^2 - 1) \end{pmatrix}\right].
\end{align}
Note that the choice $\tilde{f}<0$ and $\alpha = \pi/2$ is equivalent to 
$\tilde{f}>0$ and $\alpha=-\pi/2$.
Solving for the threshold chirality for negative drift and reversed Hall 
drift of the tracer amounts for solving a third-order polynomial inequality. 
We, therefore, restrict ourselves to give relations for the state diagram. 

Negative tracer drift, $v_\parallel/(\mu_0f_t)<0$, becomes possible for 
$\kappa>\sqrt{3}$ and $\tilde{f}<\tilde{f}^\parallel_\mathrm{rev}(\kappa, \phi)$, 
or $0<\kappa<\sqrt{3}$ and $\tilde{f}>\tilde{f}^\parallel_\mathrm{rev}(\kappa, \phi)$, 
where
\begin{equation}
\label{orthogonal_tracer_drift_parallel}
    \tilde{f}^\parallel_\mathrm{rev}(\kappa, \phi) = \frac{1 + \kappa^2 - 
    2\phi(1 - 3\kappa^2)}{2\phi\kappa(3-\kappa^2)}.
\end{equation}

A reversal of the intrinsic Hall drift of the tracer, $v_\perp/(\mu_0f_t)>0$, 
becomes possible for $\kappa>1/\sqrt{3}$ and 
$\tilde{f}>\tilde{f}^\perp_\mathrm{rev}(\kappa, \phi)$, or $0<\kappa<1/\sqrt{3}$ 
and $\tilde{f}<\tilde{f}^\perp_\mathrm{rev}(\kappa, \phi)$, where
\begin{equation}
\label{orthogonal_tracer_drift_perp}
    \tilde{f}^\perp_\mathrm{rev}(\kappa, \phi) = \frac{\kappa(1 + 
    \kappa^2 - 2\phi(3 - \kappa^2))}{2\phi(3\kappa^2-1)}.
\end{equation}
Eqs.~\eqref{orthogonal_tracer_drift_parallel} and 
\eqref{orthogonal_tracer_drift_perp} are plotted as the state diagram for 
different densities $\phi=0.04, 0.06, 0.08$ and as a function of $\kappa$ in 
Fig.~2 of the main manuscript. 

\subsection{Nonchiral tracer in chiral bath}
\label{sec:nonchiral}

The effective drift-diffusion equation \eqref{effective_FP_equation} for the 
interaction-corrected dynamics of a tracer in a Brownian fluid can be analyzed 
for various setups. We further demonstrate a common choice in the literature: 
the nonchiral tracer in a chiral bath \cite{reichhardt2019active, 
yang2021topologically, reichhardt2022activerheology, poggioli2023odd}.
 This situation is modeled as $\boldsymbol{\mu}_t=\mu_0\mathbf{1}$ 
and $\boldsymbol{\mu}_h=\mu_0(\mathbf{1} + \kappa_h\boldsymbol{\epsilon})$, 
as well as $\boldsymbol{D}_t=D_0\mathbf{1}$ and $\boldsymbol{D}_h=D_0(\mathbf{1} 
+ \kappa_h\boldsymbol{\epsilon})$. We again consider $\mathbf{f}_t \parallel 
\mathbf{f}_h \propto \hat{\mathbf{e}}_x$ for demonstration purposes. For 
simplicity we chose equal bare mobility and bare diffusion among tracer and 
host particles, different choices are straightforward. For the exemplary setup 
of parallel drift forces, the effective tracer drift of 
Eq.~\eqref{general_tracer_drift} specifies to
\begin{align}
    \begin{pmatrix}v_\parallel\\ v_\perp \end{pmatrix} &= 
\mu_0 f_t \left[ \begin{pmatrix}1 \\ 0\end{pmatrix} 
 - \frac{4\phi}{4 + \kappa_h^2} \begin{pmatrix}
     2-\tilde{f}(2 -\kappa_h^2)\\ \kappa_h(3\tilde{f} - 1) \end{pmatrix}\right].
\end{align}
Notably, and in accordance with the literature, the nonchiral tracer adopts the 
characteristic Hall drift via interactions with the chiral host medium, $v_\perp 
\propto -\phi\, \kappa_h$.

A similar analysis as before can be performed and we see negative tracer drift, 
$v_\parallel/(\mu_0f_t)<0$, for $\tilde{f}>1/(4\phi)$ and $\kappa_h > 
\kappa_{\mathrm{rev},h}^{\parallel}(\tilde{f}, \phi)$, or $\tilde{f}<1 - 1/(2\phi)$ 
and $0<\kappa_h < \kappa_{\mathrm{rev},h}^{\parallel}(\tilde{f}, \phi)$, where
\begin{equation}
\kappa_{\mathrm{rev},h}^{\parallel}(\tilde{f}, \phi) =  2\sqrt{\frac{2\phi(1 - 
\tilde{f}) -1}{1 - 4\phi\tilde{f}}}. 
\end{equation}
The condition for the reversal of (interaction-induced) Hall-drift, 
$v_\perp/(\mu_0f_t)>0$, reads $\tilde{f}>1/3$.

We leave further analysis and validation of these predictions for future work.

\subsection{Nonchiral tracer in nonchiral bath}
\label{sec:nonchiral_nonchiral}

We finally provide the analytical predictions for the nonchiral tracer in 
the nonchiral fluid, i.e., the classical reference situation. Here $\kappa_t = 
\kappa_h = 0$ in particular and we consider the situation where 
$\boldsymbol{\mu}_t=\boldsymbol{\mu}_h=\mu_0\mathbf{1}$ and $\boldsymbol{D}_h=
\boldsymbol{D}_t=D_0\mathbf{1}$. Again, we chose equal bare mobility and 
diffusion among tracer and host particles for simplicity, different choices are 
straightforward. The effective 
tracer drift of Eq.~\eqref{general_tracer_drift} specifies to
\begin{equation}
\label{tracer_drift_nonchiral_nonchiral}
\begin{pmatrix}v_\parallel\\ v_\perp \end{pmatrix} = 
\mu_0 f_t \left[ \begin{pmatrix}1 \\0\end{pmatrix} 
 - 2\phi \begin{pmatrix}
     1 - \tilde{f}\mathrm{cos}(\alpha) \\ - \tilde{f}\mathrm{sin}(\alpha)
 \end{pmatrix}
\right].
\end{equation}
In the absence of chirality, both for the tracer and the host particles, 
there is no Hall drift, $v_\perp|_{\phi=0}=0$. However, for $\alpha \neq 0$ and 
the host particles being driven in a different direction than the tracer particle,
interparticle interactions transfer an orthogonal drift $\propto \phi 
\mathrm{\sin}(\alpha)$ onto the tracer, i.e., the projected component of the 
non-aligned host drift. The tracer further shows an intuitive 
drift behavior in $v_\parallel$; for $f_t<f_h$ the tracer drift gets enhanced 
by interactions, $v_\parallel > v_\parallel|_{\phi=0}$, as the host crowd is 
subjected to a stronger drift. For $f_t>f_h$, on the contrary, collisions with 
the slower host species impede the tracer drift an $v_\parallel < 
v_\parallel|_{\phi=0}$. There is, however, a formal situation of negative tracer 
drift for $\tilde{f}<1 - 1/(2\phi)<0$ in $\phi\in[0, 0.5)$. This situation 
corresponds to the tracer particle being driven in the opposite direction of 
the host particles. If the tracer is driven with $f_t< -f_h\, 2\phi/(2\phi - 1)$, 
it still follows the opposite host-particle stream and thus, formally, responds 
negatively on its applied drift. 

\section{Simulation Details}
The simulations for this system are done in two dimensions. The Langevin 
equations for disk $i \in\{1, \ldots, N\}$ used for the simulations are
\begin{align}
    \label{eqn:lagevinx}
    \frac{\partial \mathbf{r}_i}{\partial t} &= \bm{v}_i,\\
    \label{eqn:lagevinv}
    \frac{\partial \mathbf{v}_i}{\partial t} &= -\frac{\gamma_i}{m_i}(\bm{1} - 
    \kappa_i\bm{\epsilon})\bm{v}_i + \frac{1}{m_i}\mathbf{F}_i + \frac{1}{m_i}
    \bm{\zeta}_i.
\end{align}
Here $\gamma_i$, $m_i$, and $\kappa_i$ are the friction coefficient, mass and 
oddness of particle $i$, $\bm{\epsilon}$ is the two-dimensional Levi-Civita 
symbol and $\bm{\zeta}_i$ is a zero-mean Gaussian white noise accounting for 
the fluctuations $\langle \bm{\zeta}_i(t) \bm{\zeta}^\mathrm{T}_j(t')\rangle =
2\gamma_iT\bm{1}\delta_{ij}\delta(t-t')$, where the normalized temperature $T$ 
is such that the the Boltzmann constant is unity. $\mathbf{F}_i$ is the total 
force experienced by particle $i$ and given by
\begin{equation}
    \mathbf{F}_i = \mathbf{f}_i + \sum_{j=1, i\neq j}^N \mathbf{f}_{ij},
\end{equation}
where $\mathbf{f}_i$ is the external force acting on the disk: 
$\mathbf{f}_i = \mathbf{f}_h$ if particle $i$ belong to the host species and 
$\mathbf{f}_i = \mathbf{f}_t$ for the tracer particle. $\mathbf{f}_{ij}$ is 
the force acting on particle $i$ due to interaction with particle $j$ and 
is given by
\begin{equation}
    \mathbf{f}_{ij} = \begin{cases}
        \varepsilon \left(\frac{\sigma}{|\mathbf{r}_{ij}|} \right)^{\alpha} 
        \mathbf{r}_{ij}, \ &\rm{if}\ |\mathbf{r}_{ij}|<\sigma_c\\
        0, &\rm{otherwise,}
    \end{cases} 
\end{equation}
where $\mathbf{r}_{ij} = \mathbf{r}_i - \mathbf{r}_j$, $\sigma$ is the 
diameter of the particles, $\varepsilon$ is the energy scale of the potential, 
$\alpha$ and $\sigma_c$ are the steepness and the range of the potential, 
respectively.
\label{sec:simulation}

\subsection{Parameters}
The simulations were carried out in a square box of length $L$, which was
determined by the volume fraction $\phi$ as $L = \sqrt{\pi (N-1)/(4 \phi)}$.
The total number of particles was set to $N=100$, with $N-1$ host particles
and one tracer particle. The temperature of the system is set to $T=1$. The
simulations were done with all particles having the same friction coefficient
$\gamma$ and oddness $\kappa$.  The friction coefficient is $\gamma =
\gamma_0/(1+\kappa^2)$ where $\gamma_0=1$. This normalization is done to keep
the norm of the friction matrix $\gamma(\bm{1}- \kappa\bm{\epsilon})$ unchanged
with oddness and prevent energy dissipation. The mass of the particles is
$m = m_0/(1+\kappa^2)$ and $m_0$ was chosen to be $10^{-2}$ such that
$m/\gamma = m_0/\gamma_0 = 10^{-2}$. The parameters to mimic the hard potential
were set to $\sigma = 1$, $\sigma_c = 1.01$, $\varepsilon = 100$, and
$\alpha = 17$. The external forces in the system were set to $\mathbf{f}_t  =
f_t \hat{\mathbf{e}}_x$ and $\mathbf{f}_h  = f_h \hat{\mathbf{e}}_y$ in
Figs. 1(a) and (c) in the main text and $\mathbf{f}_t  = f_t \hat{\mathbf{e}}_x$
and $\mathbf{f}_h  = -f_h \hat{\mathbf{e}}_y$ in Fig. 2(a) in the main text.
The $f_t$ values were specified and $f_h$ was calculated based on the desired
force ratio $\tilde{f}= f_h/f_t$. We used $f_t = 0.05$ for Figs. 1(c) and 2(b), 
and $f_t = 0.5$ for Fig. 1(a) owing to the smaller value of $\tilde{f}$ there.

The stochastic equations \eqref{eqn:lagevinx} and \eqref{eqn:lagevinv} were
numerically integrated using the forward Euler method \cite{kalz2022collisions}
with the timestep $\Delta t = 10^{-5}$. The trajectory of the tracer was studied
by tracking the particle's position every $10^5$ timesteps. The drift velocity 
components of the tracer were then obtained by fitting a straight line in the $x$ 
vs. $t$ and $y$ vs. $t$ data, respectively, and calculating the slopes. 
We performed 100 realizations
of such simulations with trajectory length of $t=10^3$ with different random 
numbers and show the means of these 100 slopes as the data points in the plots.